\documentclass[showpacs,aps,prl,twocolumn,superscriptaddress]{revtex4}
\usepackage{graphicx}
\usepackage{epsfig}
\usepackage{bm,hyperref}
\usepackage{CJK}
\usepackage{amsmath}

\begin{document}

\title{Bright solitons in spin-orbit-coupled Bose-Einstein condensates}
\begin{CJK*}{UTF8}{gbsn}
\author{Yong Xu(徐勇)}
\affiliation{Institute of Physics, Chinese Academy of Sciences,
Beijing 100190, China}
\affiliation{International Center for Quantum Materials, Peking
University, Beijing 100871, China}
\author{Yongping Zhang}
\affiliation{The University of Queensland, School of Mathematics and Physics,
St. Lucia, Queensland 4072, Australia}
\author{Biao Wu(吴飙)}
\email{wubiao@pku.edu.cn}
\affiliation{International Center for Quantum Materials, Peking
University, Beijing 100871, China}
\date{\today}
%
\begin{abstract}
We study bright solitons in a Bose-Einstein condensate with a
spin-orbit coupling that has been realized experimentally. Both
stationary bright solitons and moving bright solitons are found.
The stationary bright solitons are the ground states and possess
well-defined spin-parity, a symmetry involving both spatial and spin degrees
of freedom; these solitons are real valued but not positive definite, and the
number of their nodes depends on the strength of spin-orbit coupling.
For the moving bright solitons, their shapes are found to change with velocity
due to the lack of Galilean invariance in the system.
\end{abstract}


\maketitle
\end{CJK*}

\section{introduction}
Solitons are one of the most interesting topics in nonlinear systems.
The most fascinating and well-known feature of this localized
wave packet is that it can propagate without changing its shape
as a result of the balance between nonlinearity and dispersion \cite{Agrawal}.
The achievement of Bose-Einstein condensation in a dilute atomic gas
has offered a clean and parameter-controllable platform to study the properties
of solitons \cite{Kevrekidis09}. In a Bose-Einstein condensate (BEC), the nonlinearity
originates from the atomic interactions and is manifested by
the nonlinear term in the Gross-Pitaevskii equation (GPE), which
is the  mean-field description of BEC \cite{Pethick}.
With attractive and repulsive interatomic interactions, the GPE can have bright
and dark solitons solutions, respectively.
Such dark and bright solitons in BECs have been studied extensively
both theoretically \cite{Carr,Jackson,Dum,Herrero,Muryshev,Fedichev,BWu} and
experimentally \cite{Anderson1,Burger,Denschlag,Khaykovich, Strecker,Cornish,Chin,Eiermann}.

The developments with two-component BECs have further enriched the investigation of
solitons in matter waves. The two-component BECs not only introduce more tunable
parameters, for example, the interaction between the two species, but
also bring in novel nonlinear structures which have no counterparts
in the scalar BEC, such as dark-bright solitons (one component is a
dark soliton while the other is bright) \cite{Busch,Kevrekidis05,Becker,Hamner},
dark-dark solitons \cite{Ohberg},
bright-bright solitons \cite{Victor,Adhikari,Salasnich}, and
domain walls \cite{Kasa,Kevrekidis04,Jacopo}.

Recently, in a landmark experiment, the Spielman group at NIST have engineered
a synthetic spin-orbit coupling (SOC) for a BEC \cite{Lin}. In the experiment, two
Raman laser beams are used to couple a two-component BEC. The
momentum transfer between lasers and atoms leads to synthetic
spin-orbit coupling
\cite{Ruseckas,Zhu,Liu,Sranescu,Dalibard,Sau,Campbell,Juzeliunas,Zhang,Spielman}.
This kind of spin-orbit coupling has subsequently been realized for neutral atoms
in other laboratories \cite{Bloch11,JZhang,Zwierlein,Pan}.
These experimental breakthroughs \cite{Lin,Fu,Pan} have stimulated extensive theoretical
investigation of the properties of spin-orbit-coupled BECs \cite{Zhai,Ho,Wucongjun,Yongping,Yip,Xu,
Kawakami,Xiaoqiang,Zhou,Radic,Sinha,Hu,Ramach,Ozawa,Yun,Qizhong,Bijl,Larson,
Xiaoqiang2,Zheng,Yun2,Yongping2}, which includes some early studies
on solitons.
For example, bright-soliton solutions were found analytically for
spin-orbit-coupled BECs by neglecting the kinetic energy \cite{Merkl}.
Dark solitons for such a system were studied in a one-dimensional ring \cite{Fialko}.
In this work we conduct a systematic study of bright solitons for a
BEC with attractive interactions and the experimentally realized
SOC \cite{Lin,JZhang,Zwierlein,Pan}. By solving the GPE both analytically
and numerically, we find that these solitons possess a number of novel
properties due to the SOC.

In particular, we find that the stationary bright solitons that are
the ground state of the system have nodes in their wave function.
For a conventional BEC without SOC, its ground state must be nodeless thanks to the
``no-node" theorem for the ground state of a bosonic system
\cite{Wucongjun2}. Furthermore, these solitons are found to have
well-defined spin-parity, a symmetry that involves both spatial and spin
degrees of freedom, and can exist in systems with SOC.

We have also found solutions for moving bright solitons. They have the very interesting
feature that their shapes change dramatically with increasing velocity. For a
conventional BEC, the shape of a soliton does not change with velocity due to the
Galilean invariance of the system.
In other nonlinear systems, such as Korteweg-de Vries KdV systems, the shape of a soliton
changes only in height and width with velocity \cite{Newell}. In stark contrast,
bright solitons in a BEC with SOC can change shape dramatically
from nodeless to having many nodes with varying velocity. The new feature
arises because of the lack of Galilean invariance due to SOC \cite{Qizhong}.
It is worthwhile to note that a similar model was proposed a long time ago
in the context of nonlinear birefringent fibers ~\cite{Malomed1991}.
This shows that our results  will find applications in nonlinear optics.

\section{Model equation}

A BEC with the experimentally realized SOC is described by the following GPE
\begin{equation}
i\hbar\frac{\partial\Psi}{\partial
t}=\Big[\frac{1}{2m}(p_x+\hbar\kappa{\sigma_y})^{2}+
\hbar\Delta\sigma_{z}-g{\Psi}^{\dagger}\cdot{\Psi}\Big]{\Psi}\,,
\label{gpe}
\end{equation}
where the spinor wavefunction ${\Psi}=(\Psi_1,\Psi_2)^T$, and
${\Psi}^{\dagger}\cdot{\Psi}=|\Psi_1|^2+|\Psi_2|^2$ with $\Psi_1$
for up-spin and $\Psi_2$ for down-spin. The nonlinear
coefficient $-g<0$ is for attractive interatomic interactions, and
we have taken $g_{11}=g_{22}=g_{12}$ for simplicity. The
SOC is realized experimentally by two
counter-propagating Raman lasers that couple two hyperfine ground states
$\Psi_1$ and $\Psi_2$. The strength of SOC $\kappa$
depends on the relative incident angle of the Raman beams and can be
changed \cite{Yongping2}. The Rabi frequency $\Delta$ can be tuned
easily by modifying the intensity of the Raman beams.
$\sigma$ are Pauli matrices. A bias homogeneous magnetic field
is applied along the $y$ direction. We consider the case that
the radial trapping frequency is large and, therefore, the system is effectively
one dimensional \cite{Khaykovich, Strecker}.

For numerical simulation, we rewrite Eq. (\ref{gpe})
in a dimensionless form by scaling energy with $\hbar\Delta$ and
length with $\sqrt{\hbar/m\Delta}$. The dimensionless GPE is
\begin{equation}
\label{t_GP}
i\partial_{t}{\Phi}=\big[-\frac{1}{2}\partial_{x}^{2}+i\alpha\sigma_{y}\partial_{x}+\sigma_{z}-\gamma{\Phi}^{\dagger}\cdot{\Phi}\big]{\Phi}\,.
\end{equation}
The dimensionless parameters $\alpha=-\kappa\sqrt{\hbar\Delta /m}$
and $\gamma=Ng\sqrt{m/(\hbar\Delta)}/\hbar$ with $N$ being the total
number of atoms. The dimensionless wavefunctions $\Phi$ satisfy $\int dx
(|\Phi_1|^2+|\Phi_2|^2)=1$. The SOC term $i\alpha\sigma_{y}\partial_{x}$ in  Eq. (\ref{t_GP})
indicates that spin $\sigma_{y}$ only couples the momentum in the $x$ direction. The
energy functional of our system is
\begin{align}
\label{eng}
E=\int dx
\Big[&\frac{1}{2}|\partial_x\Phi_1|^2+\frac{1}{2}|\partial_x\Phi_2|^2+|\Phi_1|^2-|\Phi_2|^2\notag
\\
&+\alpha\Phi_1^*\partial_x\Phi_2-\alpha\Phi_2^*\partial_x\Phi_1\\&-\frac{\gamma}{2}(|\Phi_1|^4
+|\Phi_2|^4+2|\Phi_1|^2|\Phi_2|^2)\Big]\notag.
\end{align}

\begin{figure}[t]
\begin{center}
\scalebox{0.45}[0.45]{\includegraphics[20,335][602,589]{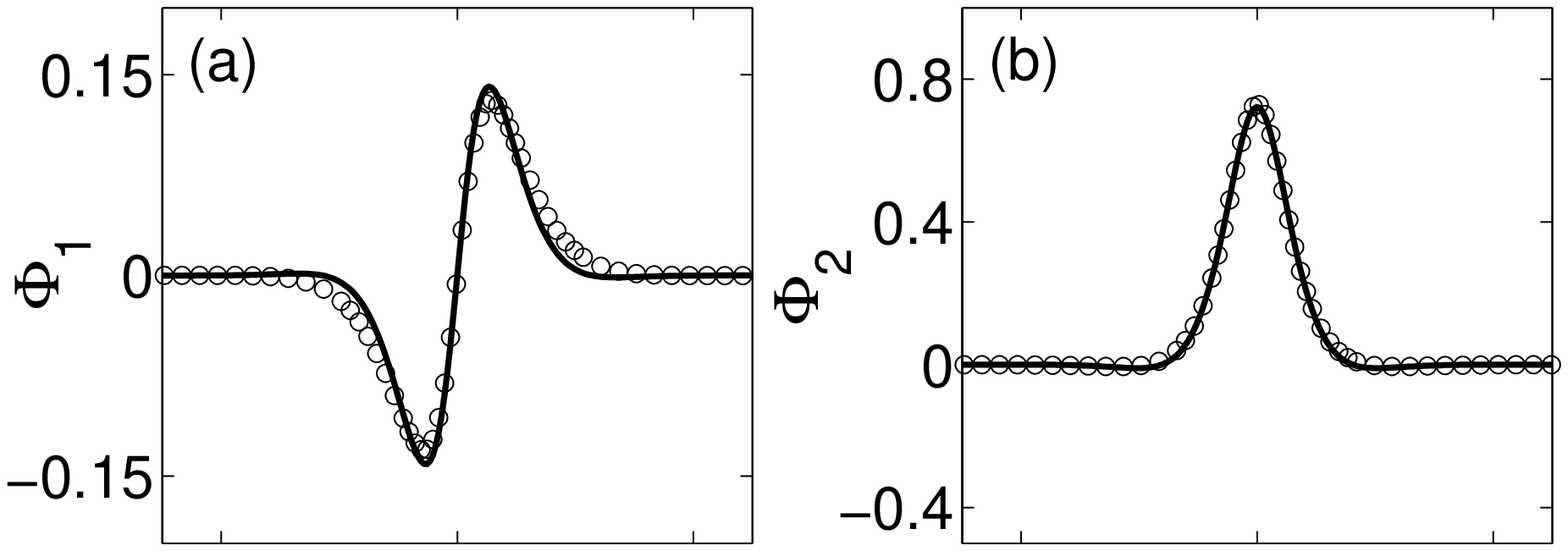}}
\scalebox{0.45}[0.45]{\includegraphics[20,335][602,556]{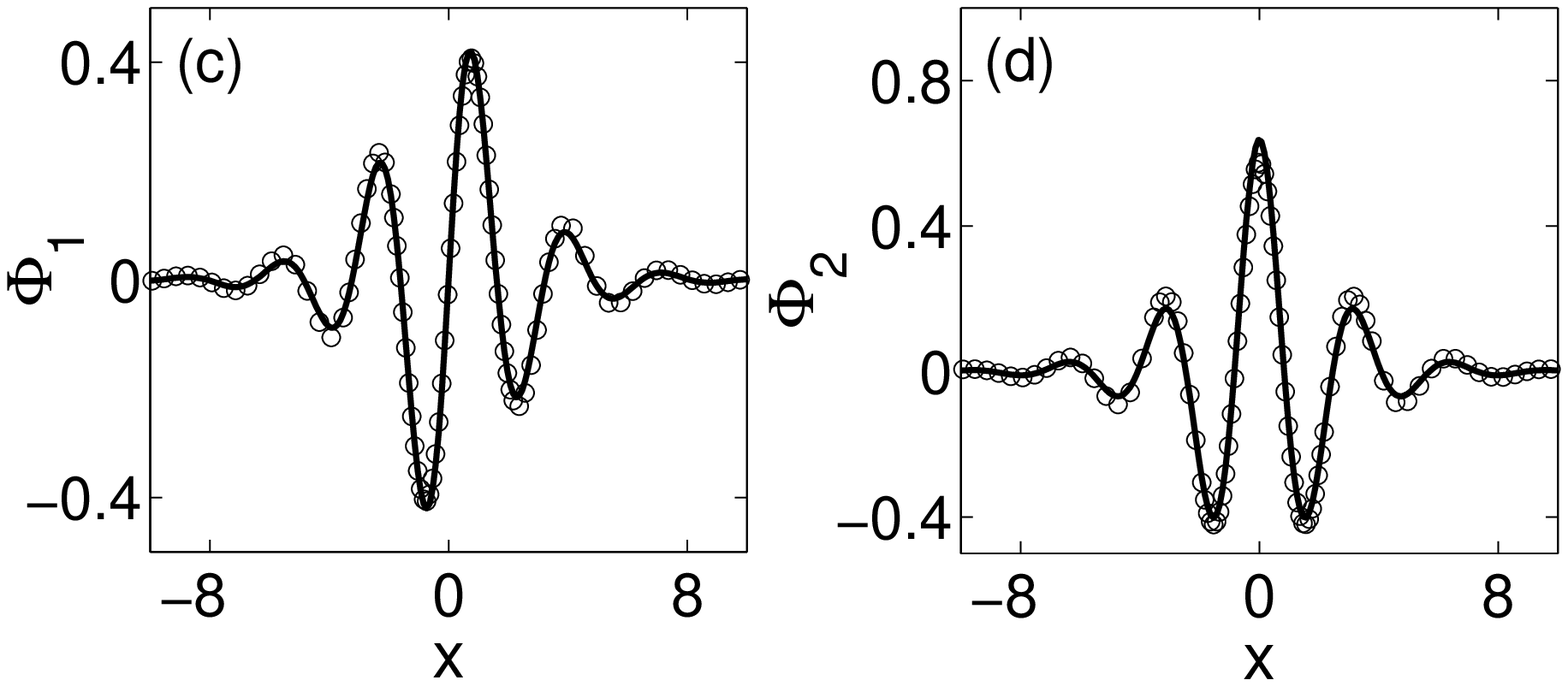}}
\caption{Stationary bright solitons at $\gamma=1.0$. The
solid lines are numerical results and the circles are
from the variational method. In (a) and (b), $\alpha=1.0$.   In
(c) and (d), $\alpha=2.0$.} \label{G_va_strip}
\end{center}
\end{figure}

\section{stationary bright solitons}
We focus on the simplest stationary bright solitons, which are the ground
states of the system. To find these solitons, we solve  Eq. (\ref{t_GP}) by using
an imaginary time-evolution method. Two typical bright solitons are shown in
Fig. \ref{G_va_strip}. One interesting feature is immediately noticed.
There are ``nodes" in these ground-state bright solitons. This is
very different from the conventional BEC, where there are no
nodes in this kind of ground state soliton as demanded by the
no-node theorem for the ground state of a boson system.
Our results confirm that this no-node theorem does not hold for systems
with SOC \cite{Wucongjun2}.

There can exist a unique symmetry for systems with SOC,
spin-parity, which involves both spatial and spin degrees of freedom.
The operator for spin-parity is defined as
\begin{equation}
\mathcal {P}=P\sigma_z,
\end{equation}
where $P$ is the parity operator. It is easy to verify that our system
is invariant under the action of spin-parity $\mathcal {P}$. By direct observation,
one can see that the bright solitons shown in Fig. \ref{G_va_strip} satisfy
\begin{equation}
\mathcal {P}\begin{pmatrix}\Phi_1(x)\\\Phi_2(x)
\end{pmatrix}=-\begin{pmatrix}\Phi_1(x)\\\Phi_2(x)
\end{pmatrix}\,,
\end{equation}
as the up component $\Phi_1$ has odd parity while the other component
$\Phi_2$ is even. Therefore, these bright solitons have spin-parity $-1$.
In fact, all the ground state bright solitons that we have found have
spin-parity $-1$. That the eigenvalue of $\mathcal{P}$ for these solitons
is $-1$ and not $1$ can be understood in the following manner. When the strength
of SOC $\alpha$ decreases to zero, the up component $\Psi_1$ shrinks
to zero and only the down component survives. Since the system becomes
a conventional BEC without SOC, the no-node theorem demands that
the surviving down component has even symmetry. As the SOC is turned
up continuously and slowly, the symmetry of the second component
should remain and the spin parity has to be $-1$.

\begin{figure}[!h]
\begin{center}
\scalebox{0.54}[0.54]{\includegraphics[120,289][397,650]{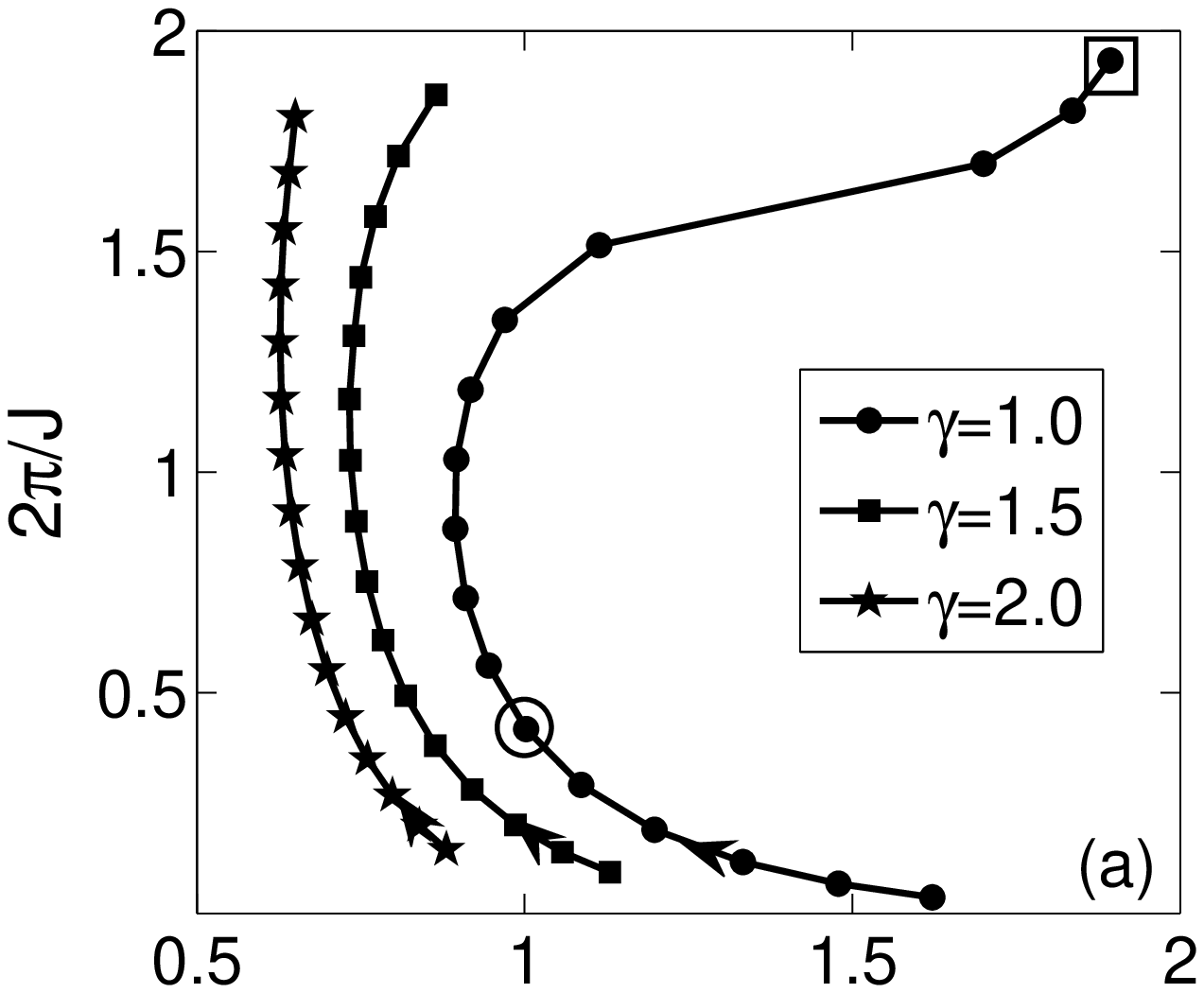}}
\scalebox{0.54}[0.54]{\includegraphics[137,250][397,590]{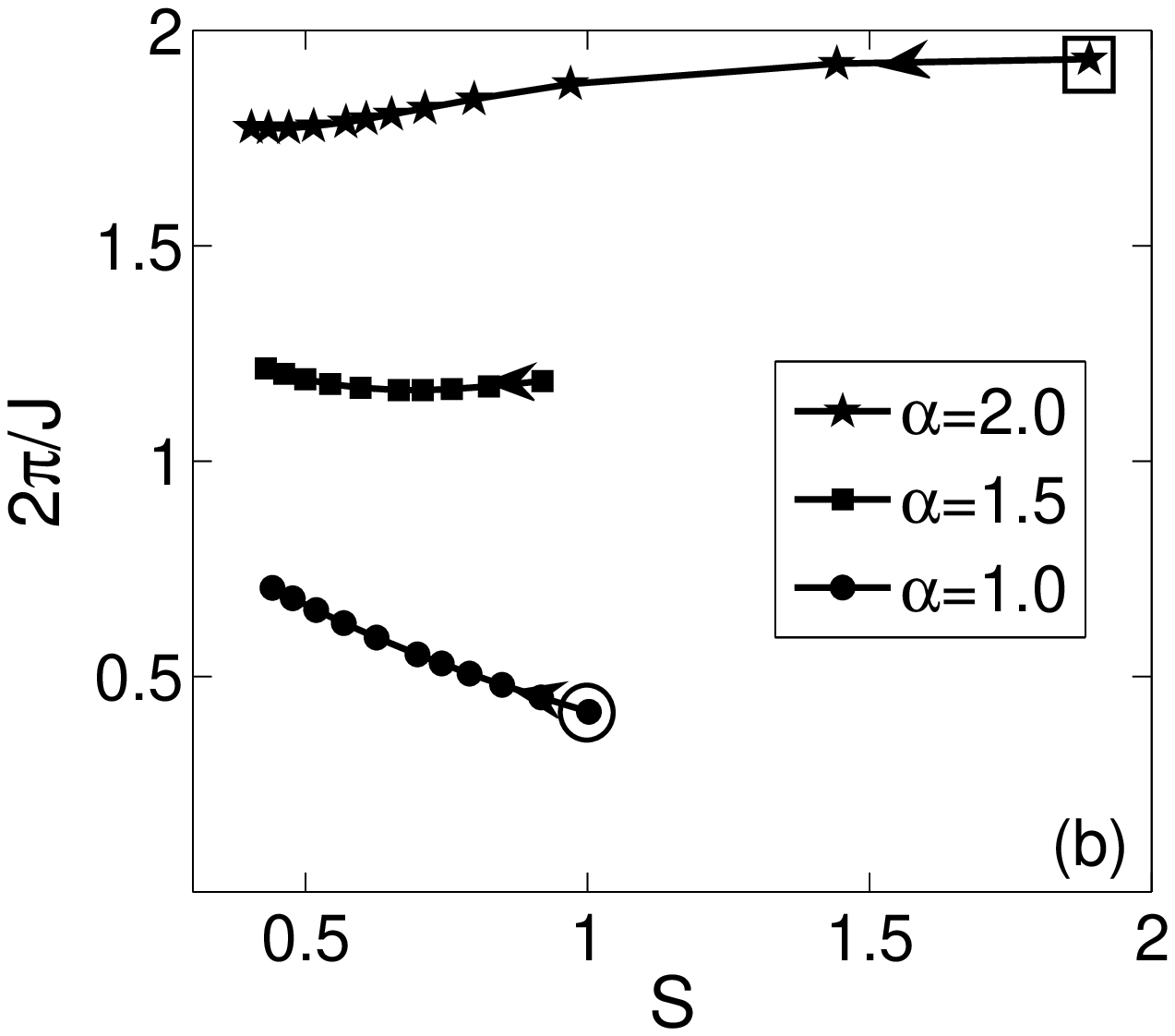}}
\caption{The relation between the number of soliton nodes $2\pi/J$ and
the soliton width $S$. (a) Circles, squares,
and stars are for $\gamma=1.0,~1.5,~2.0$, respectively.
Here $\alpha$ increases from $0.5$ to $2.0$ (the arrow direction). (b)
Circles, squares, and stars are for $\alpha=1.0,~1.5,~2.0$.
Here $\gamma$ increases from $1.0$ to $4.0$ (the arrow direction).
The open circle and square correspond to the bright solitons in
Figs. \ref{G_va_strip}(a), and \ref{G_va_strip}(b) and
Figs. \ref{G_va_strip}(c), and \ref{G_va_strip}(d), respectively. }
\label{J_sigma}
\end{center}
\end{figure}

For a more detailed analysis of these bright solitons, we attempt to find
an analytical approximation for the wavefunctions using the variational method.
Motivated by the features of the stationary bright solitons shown
in Fig. \ref{G_va_strip}, we propose the following trial wave functions
for these solitons:
\begin{eqnarray}
{\Phi}=
\begin{pmatrix}A\sin(2\pi x/J) \\ B\cos(2\pi x/J)\end{pmatrix}\text{sech}(x/S)\,.
\label{ansatz}
\end{eqnarray}
The  parameters $A$, $B$, $J$, and $S$ are determined by minimizing
the energy functional in Eq. (\ref{eng}) with the constraining normalization.
The results of the trial wave functions are compared with the numerical results
in Fig. \ref{G_va_strip}, where it can be seen that they are in excellent agreement.

It is clear from Eq. (\ref{ansatz}) that the parameter $2\pi/J$
can be regarded roughly as the number of nodes in the
bright solitons, while $S$ is for the overall width of the soliton.
Both of them depend on the SOC strength $\alpha$ and the interaction
strength $\gamma$. In Fig. \ref{J_sigma} we have plotted the relation
between $2\pi/J$ and $S$, demonstrating how the number of nodes
is related to the soliton width for different values of $\alpha$
and $\gamma$. As shown in the Fig.~\ref{J_sigma}, for solitons
with the same number of nodes, they are wider for smaller interaction
strength $\gamma$ [Fig. \ref{J_sigma}(a)]; the solitons with the same
width have more nodes for larger SOC strength $\alpha$ [Fig. \ref{J_sigma}(b)].

\begin{figure}[!h]
\begin{center}
\scalebox{0.5}[0.5]{\includegraphics[6,273][443,590]{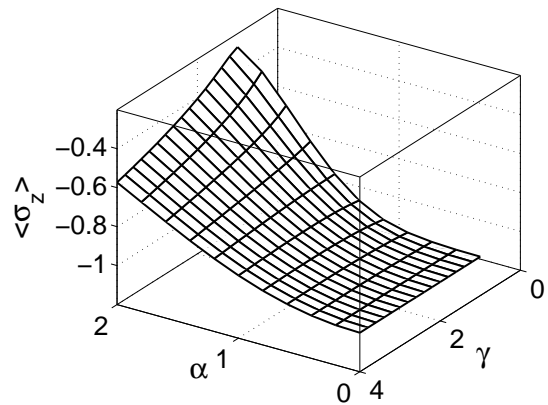}}
\caption{The spin polarization $\langle\sigma_z\rangle$ as
a function  of $\alpha$ and $\gamma$.}
\label{S_alpha}
\end{center}
\end{figure}

It has been reported that there exists a quantum phase transition
for the ground states in the spin-orbit-coupled system with repulsive
interaction \cite{Yongping2}. It is
interesting to check whether such a phase transition exists for
the case of attractive interaction. For this purpose, we have computed
the spin polarization $\langle\sigma_z\rangle=\int dx(\Phi_1^2-\Phi_2^2)$
for these bright solitons and the results are plotted
in Fig. \ref{S_alpha}. We see that for a given $\gamma$, it changes
smoothly with the SOC strength $\alpha$. For the cases of repulsive interaction
and no interaction, the spin polarization is found to change
sharply with $\alpha$, indicating a quantum phase transition \cite{Yongping2}.
The smooth behavior of Fig. \ref{S_alpha} suggests there is no quantum phase
transition.

\section{ Moving bright solitons}
After the study of stationary bright solitons, we turn our attention
to moving bright solitons. For a conventional BEC without SOC,
it is straightforward to find a moving bright soliton from a stationary
soliton: if the wave function $\Phi_s$ describes a stationary soliton,
then $\exp(ivx)\Phi_s(x-vt)$ is the wave function, up to a trivial phase,
for a soliton moving at speed $v$. This is due to the invariance of
the system under Galilean transformations.

However, Galilean invariance is violated for a spin-orbit-coupled
BEC \cite{Qizhong}. To see this explicitly,
we assume moving solitons having the following form:
\begin{equation}
{\Phi_M}(x,t)=\Phi_v(x-v t, t)\exp({ivx-i\frac{1}{2} v^{2}t}),
\end{equation}
where $\Phi_v$ is a localized function. Substitution of ${\Phi_M}(x,t) $
into Eq. (\ref{t_GP}) yields
\begin{equation}
\label{t_GP_v}
 i\partial_{t}\Phi_v=\big[-\frac{1}{2}\partial_{x}^{2}+
 \alpha\sigma_{y}(i\partial_{x}-v)+\sigma_{z}-\gamma\Phi^{\dagger}_v
 \cdot {\Phi_v}\big]\Phi_v\,.
\end{equation}
Compared to Eq. (\ref{t_GP}), this dynamical equation has
an additional term $\alpha v\sigma_{y}$, indicating the
violation of Galilean invariance. This violation means that it is no
longer a trivial task to find a moving bright soliton for a BEC with SOC.

\begin{figure}[!ht]
\begin{center}
\scalebox{0.42}[0.42]{\includegraphics[0,324][602,589]{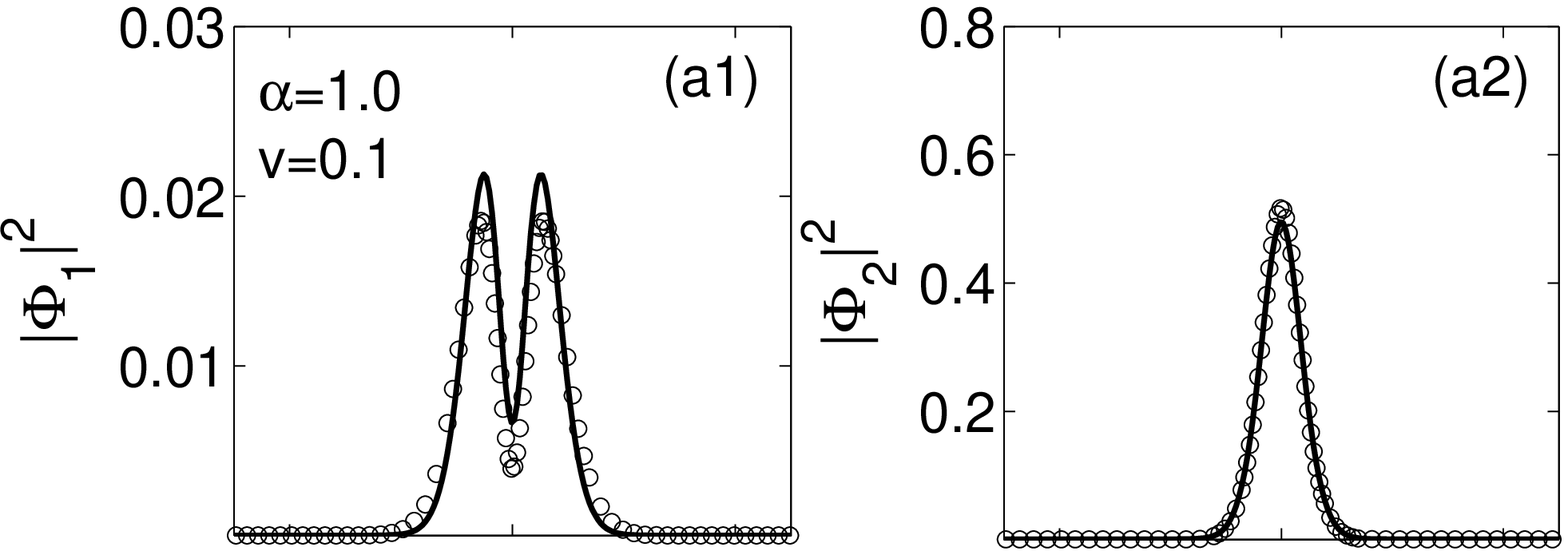}}
\scalebox{0.42}[0.42]{\includegraphics[0,324][602,540]{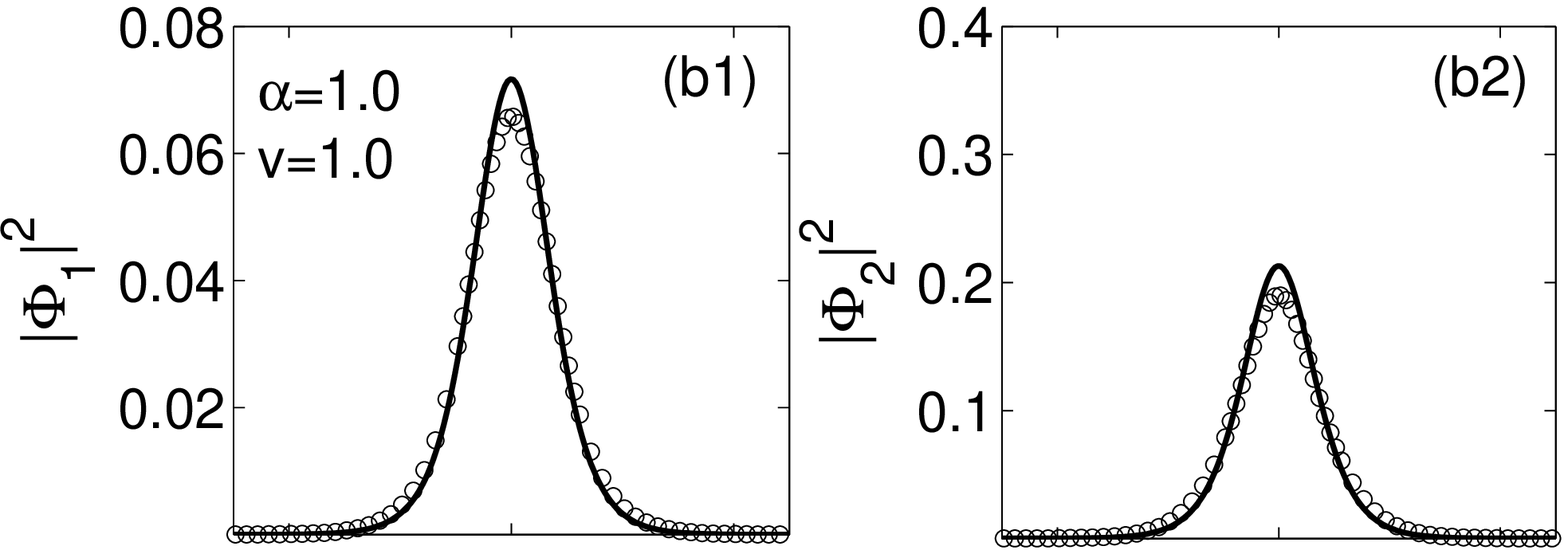}}
\scalebox{0.42}[0.42]{\includegraphics[0,324][602,540]{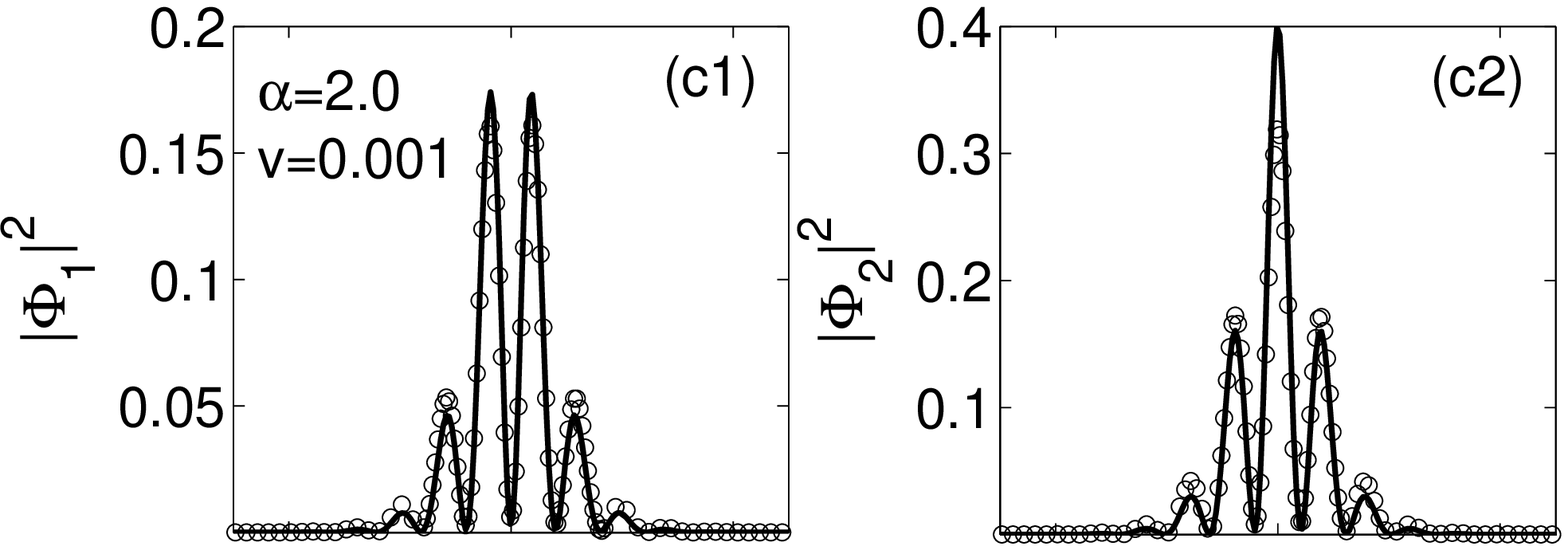}}
\scalebox{0.42}[0.42]{\includegraphics[0,324][602,540]{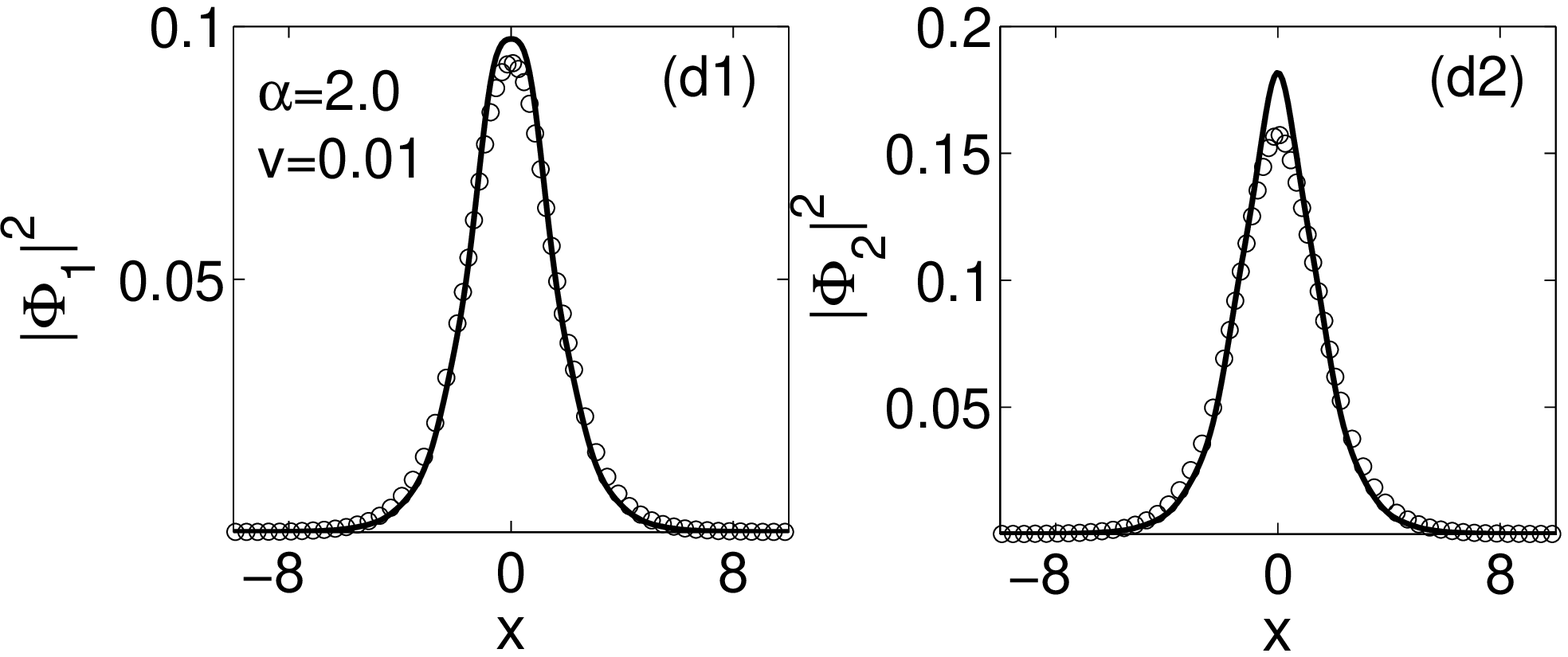}}
\caption{Moving bright soliton profiles $\Phi_v(x)$ from the
numerical calculation (solid line) and the variational method (circles)
with Eq. (\ref{ansatz2}). $\gamma=1.0$.
In (a) and (b), $\alpha=1.0$. In (c) and (d)$\alpha=2.0$.}
\label{G_va_strip_v}
\end{center}
\end{figure}

To find moving bright solitons, we numerically solve Eq. (\ref{t_GP_v})
using the imaginary time-evolution method. Two typical moving bright solitons
are shown in Fig. \ref{G_va_strip_v}, where we see clearly that
the shapes of moving bright solitons change with their velocities.
As seen in Figs. \ref{G_va_strip_v}(a) and \ref{G_va_strip_v}(b),
when the velocity $v$ is changed from 0.1 to 1, the density of the
up component changes from having two peaks to having only one.
At a larger SOC strength, such as $\alpha=2$ in Figs. \ref{G_va_strip_v}(c)
and \ref{G_va_strip_v}(d), a small change in velocity
leads to a dramatic change in the soliton profiles.

Similar to the stationary soliton, these moving bright solitons can
also be found with the variational method by minimizing the
energy functional with the following trial wave function:
\begin{equation}
\Phi_v(x)=
\begin{pmatrix}A\left[\sin\frac{2\pi x}{J}+\rho_1i\cos\frac{2\pi x}{J}\right]
\\ B\left[\cos\frac{2\pi x}{J}+\rho_2i\sin\frac{2\pi x}{J}\right]\end{pmatrix}\text{sech}\frac{x}{S},
\label{ansatz2}
\end{equation}
with two new parameters $\rho_1$ and $\rho_2$. When $\rho_1=\rho_2=0$,
we recover the stationary soliton in Eq. (\ref{ansatz}). The solutions obtained with
the variational method are plotted in Fig. \ref{G_va_strip_v} and they agree well
with the numerical results. It is clear from the trial wave function that the moving
bright soliton has no well-defined spin-parity $\mathcal {P}$.

\begin{figure}[t]
\begin{center}
\scalebox{0.38}[0.38]{\includegraphics[47,225][565,610]{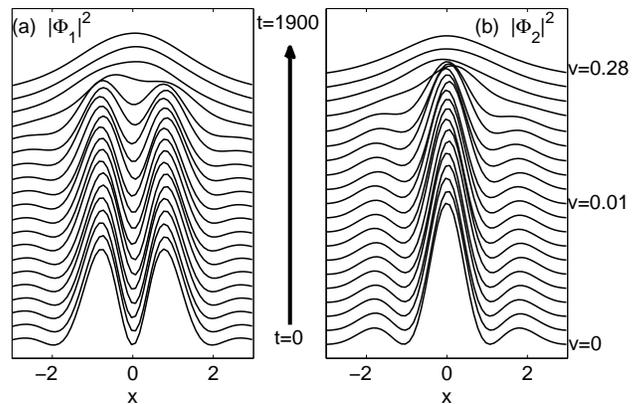}}
\caption{Dynamical evolution of a bright soliton under the influence of
a small linear potential $V(x)=\eta x$.
$\alpha=1.7$, $\gamma=1.0$, $\eta=0.001$. For better comparison of shapes, the centers
of the soliton densities have all been shifted to the zero point. The velocity of the
soliton is labeled at the right of (b).}
\label{linear}
\end{center}
\end{figure}

These moving bright solitons are adiabatically linked to the stationary bright solitons.
To see this, we slowly accelerate the stationary bright soliton by
adding a small linear potential in Eq. (\ref{t_GP}) integrating with a stationary bright soliton
as the initial condition. The dynamical evolution of this soliton is shown in Fig. \ref{linear},
where we see clearly how a stationary soliton is developed into a moving soliton with
its shape changing constantly. Note that the centers of solitons in Fig. \ref{linear}
have all been shifted to the zero point for better comparison between shapes.

We note that there are other solitons, which also change their shapes with velocity,
for example, dark solitons in a BEC and bright solitons in the KdV system \cite{Newell}.
This change is also caused by the lack of the Galilean invariance in the system. For the dark soliton,
the constant background provides a preferred reference frame and breaks the
Galilean invariance. In the KdV system, the violation is caused by the non-quadratic
linear  dispersion. However, in these systems, the change in shape with velocity is
not as dramatic: there are only changes in the height and width of the solitons.
With a spin-orbit-coupled BEC, the number of peaks in the solitons can change with a slight
change in velocity.

\section{Conclusion}

We have systematically studied both stationary
and moving bright solitons in a spin-orbit-coupled BEC.
These bright solitons have features not present without
spin-orbit coupling, for example, the existence of nodes and spin-parity
in the stationary bright solitons, and the change in shape with velocity
in the moving bright solitons. Although there are multiple peaks in
the soliton profiles, the bright solitons that we have found are single solitons.
It would be very interesting to seek out multiple-solitons solutions for this
spin-orbit-coupled system. These bright solitons should be able to
be observed in experiment. One can apply the same Raman laser setup
to generate the synthetic spin-orbit coupling for an optical dipole-trapped
$^7$Li condensate where the interatomic interaction is
attractive by nature.

\section{acknowledgments}
We would like to thank Matthew Davis for critical readings and
insightful comments. Y.X. thanks Y. Hu and M. Gong for very helpful discussions.
Y.X. and B.W. are supported by the NBRP of China (Grants No. 2012CB921300 and No. 2013CB921900),
the NSF of China (Grants. No. 10825417 and No. 11274024), the RFDP of China (Grant No. 20110001110091).

\emph{Note added}. Recently, we noticed a newly posted preprint
on bright solitons in a BEC with SOC \cite{Achilleos12}.
Some of their results overlap with ours.

\end{document}